\documentclass{article}

    \usepackage[preprint]{neurips_2025}

\usepackage[utf8]{inputenc} 
\usepackage[T1]{fontenc}    
\usepackage{hyperref}       
\usepackage{url}            
\usepackage{booktabs}       
\usepackage{amsfonts}       
\usepackage{nicefrac}       
\usepackage{microtype}      
\usepackage{xcolor}         
\usepackage{amsmath}
\usepackage{graphicx}
\usepackage{caption}
\usepackage{subcaption}
\usepackage{algpseudocode}
\usepackage{algpseudocode}
\usepackage[ruled,vlined]{algorithm2e}
\usepackage[symbol]{footmisc}

\usepackage{booktabs}
\usepackage{lipsum} 

\title{FastFLUX: Pruning FLUX with \\Block-wise Replacement and Sandwich Training}

\author{
Fuhan Cai$^{1,}$\footnotemark[1]\quad Yong Guo$^{2,*}$\quad Jie Li$^{3,*}$ \quad Wenbo Li$^{4}$\quad Jian Chen$^{3}$ \quad Xiangzhong Fang$^{1}$\\
$^1$Shanghai Jiao Tong University \\
$^2$Max Planck Institute for Informatics \\
$^3$South China University of Technology \\
$^4$Chinese University of Hong Kong \\
}

\begin{document}

\footnotetext[1]{Equal Contribution.}

\maketitle

\begin{abstract}
Recent advancements in text-to-image (T2I) generation have led to the emergence of highly expressive models such as diffusion transformers (DiTs), exemplified by FLUX. However, their massive parameter sizes lead to slow inference, high memory usage, and poor deployability. Existing acceleration methods (e.g., single-step distillation and attention pruning) often suffer from significant performance degradation and incur substantial training costs. To address these limitations, we propose \textbf{FastFLUX}, an architecture-level pruning framework designed to enhance the inference efficiency of FLUX. At its core is the \textbf{Block-wise Replacement with Linear Layers (BRLL)} method, which replaces structurally complex residual branches in ResBlocks with lightweight linear layers while preserving the original shortcut connections for stability. Furthermore, we introduce \textbf{Sandwich Training (ST)}, a localized fine-tuning strategy that leverages LoRA to supervise neighboring blocks, mitigating performance drops caused by structural replacement. Experiments show that our FastFLUX maintains high image quality under both qualitative and quantitative evaluations, while significantly improving inference speed, even with 20\% of the hierarchy pruned. Our code will be available soon.

\end{abstract}

\section{Introduction}
Recent advances in diffusion-based generative models~\cite{zhang2024attentioncalibrationdisentangledtexttoimage,esser2024scaling,podell2024sdxl,sauer2023adversarialdiffusiondistillation,chen2023pixartalphafasttrainingdiffusion,zhang2023addingconditionalcontroltexttoimage} have substantially improved the fidelity and controllability of text-to-image synthesis. Transformer-based architectures such as DiT~\cite{peebles2022dit} and FLUX~\cite{flux2024} further enhance generation quality by stacking deep hierarchical blocks that increase model expressiveness and scalability. However, these designs also introduce significant computational overhead, including large memory consumption and high inference latency, which limits their suitability for real-time generation and deployment on resource-constrained hardware.

Although prior studies have investigated model compression through pruning~\cite{castells2024ld,fang2023structural}, knowledge distillation~\cite{chen2025sana,meng2023distillation,salimans2022progressive}, and quantization~\cite{li2024svdquant,so2023temporal,he2023efficientdm}, 
these approaches primarily focus on parameter-level optimization, largely overlooking the hierarchical structural organization of deep generative transformers. 
Specifically, hierarchical redundancy in diffusion-based generative models remains largely underexplored, despite being a significant source of inefficiency in modern deep transformer architectures.
Furthermore, many existing techniques~\cite{Shirkavand2025,fang2023structural} require full-model fine-tuning after pruning or incur substantial performance degradation, which limits their practical applicability.

In this work, we revisit the problem of hierarchical redundancy in large diffusion transformers and propose \textbf{FastFLUX}, a general and efficient framework for architecture-level compression.
FastFLUX operates by selectively replacing the computationally heavy residual branches of FLUX blocks with lightweight linear alternatives, while preserving generation quality through a localized fine-tuning strategy.
Specifically, we introduce \textbf{Block-wise Replacement with Linear Layers (BRLL)}, which targets the high-cost residual pathways in ResBlocks. 
In FLUX, each ResBlock contains a heavy residual branch and a lightweight shortcut connection; the residual branch dominates both FLOPs and latency during inference.
BRLL substitutes this branch with a simple linear layer while preserving the shortcut path, ensuring stable information flow.
These linear layers can be initialized via least-squares estimation, providing a training-free initialization with controllable accuracy–efficiency trade-offs.
To mitigate performance degradation after replacement, we propose a \textbf{Sandwich Training (ST)} strategy. 
Instead of full-model fine-tuning, ST performs localized adaptation by applying LoRA~\cite{hu2021LoRAlowrankadaptationlarge} modules to the neighboring unpruned layers of each replaced block, forming a small “sandwich” around the linear module.
This local supervision enables the replaced branch to adapt effectively within its context, achieving stable and high-quality generation with minimal training overhead.
Using these approaches, we prune 20\% of the FLUX model while retaining high visual fidelity and significantly improving inference efficiency.

To sum up, our \textbf{contributions} are as follows:
\begin{itemize}
    \item We propose \textbf{FastFLUX}, a novel architecture-level pruning framework built upon the 12B FLUX model. FastFLUX is lightweight, efficient, and incurs low training cost. Even under a 20\% pruning ratio, it maintains high visual generation quality.
    \item We introduce \textbf{Block-wise Replacement with Linear Layers (BRLL)}, which replaces complex residual branches with linear layers. These replacements support training-free initialization via least-squares regression, with controllable performance degradation.
    \item We propose \textbf{Sandwich Training (ST)}, a localized fine-tuning strategy that applies LoRA to the adjacent unpruned layers of each replaced block to compensate for performance loss after pruning, enabling targeted adaptation with minimal training overhead.
\end{itemize}

\section{Related Work}
\textbf{Diffusion-based Generative Models}. 
Diffusion models have achieved remarkable progress across diverse domains~\cite{cao2024survey}, establishing foundations through discrete frameworks like DDPM~\cite{ho2020ddpm} and latent space formulations such as Latent Diffusion~\cite{rombach2022ldm}. Transformer-based architectures like DiT~\cite{peebles2022dit} demonstrate strong scalability, with recent models such as FLUX~\cite{flux2024} further increasing depth to enhance generation quality. However, deeper hierarchical stacks introduce substantial computational and memory overhead, motivating the need for efficient compression strategies.

\textbf{Model Quantization and Pruning}.
To reduce computational costs, diffusion models employ quantization and pruning techniques~\cite{shen2025efficient}. Post-training quantization methods like PTQ4DM~\cite{shang2023post}, Q-Diffusion~\cite{li2023q}, and PTQD~\cite{he2023ptqd} address timestep-specific calibration challenges, while quantization-aware training approaches such as TDQ~\cite{so2023temporal} and QALoRA~\cite{he2023efficientdm} learn optimized representations. Complementary pruning methods remove redundant weights via Taylor expansion~\cite{fang2023structural}, latent space guidance~\cite{castells2024ld}, or dynamic programming~\cite{kim2024layermerge}, achieving substantial size reduction with minimal quality loss.

\textbf{Transformer Compression}.
Transformer compression tackles computational challenges through multiple strategies~\cite{tang2024survey}. Knowledge distillation transfers capabilities from large teachers to compact students~\cite{sanh2019distilbert,touvron2021training}, while quantization reduces precision via methods like FQ-ViT~\cite{lin2021fq} and PTQ-ViT~\cite{liu2021post}. Pruning techniques eliminate redundancy through head removal~\cite{yu2023x} or context selection~\cite{anagnostidis2023dynamic}. Architectural innovations target attention mechanisms~\cite{ainslie2023gqa,hatamizadeh2023fastervit}, feed-forward networks~\cite{du2022glam,touvron2023llama}, or propose alternative designs like Mamba~\cite{gu2024mamba,zhu2024vision}. While these methods address parameter and structural redundancy, explicit compression of depth-wise redundancy in diffusion transformers remains underexplored.

\begin{figure}[htbp]
  \centering
  \includegraphics[width=1\linewidth]{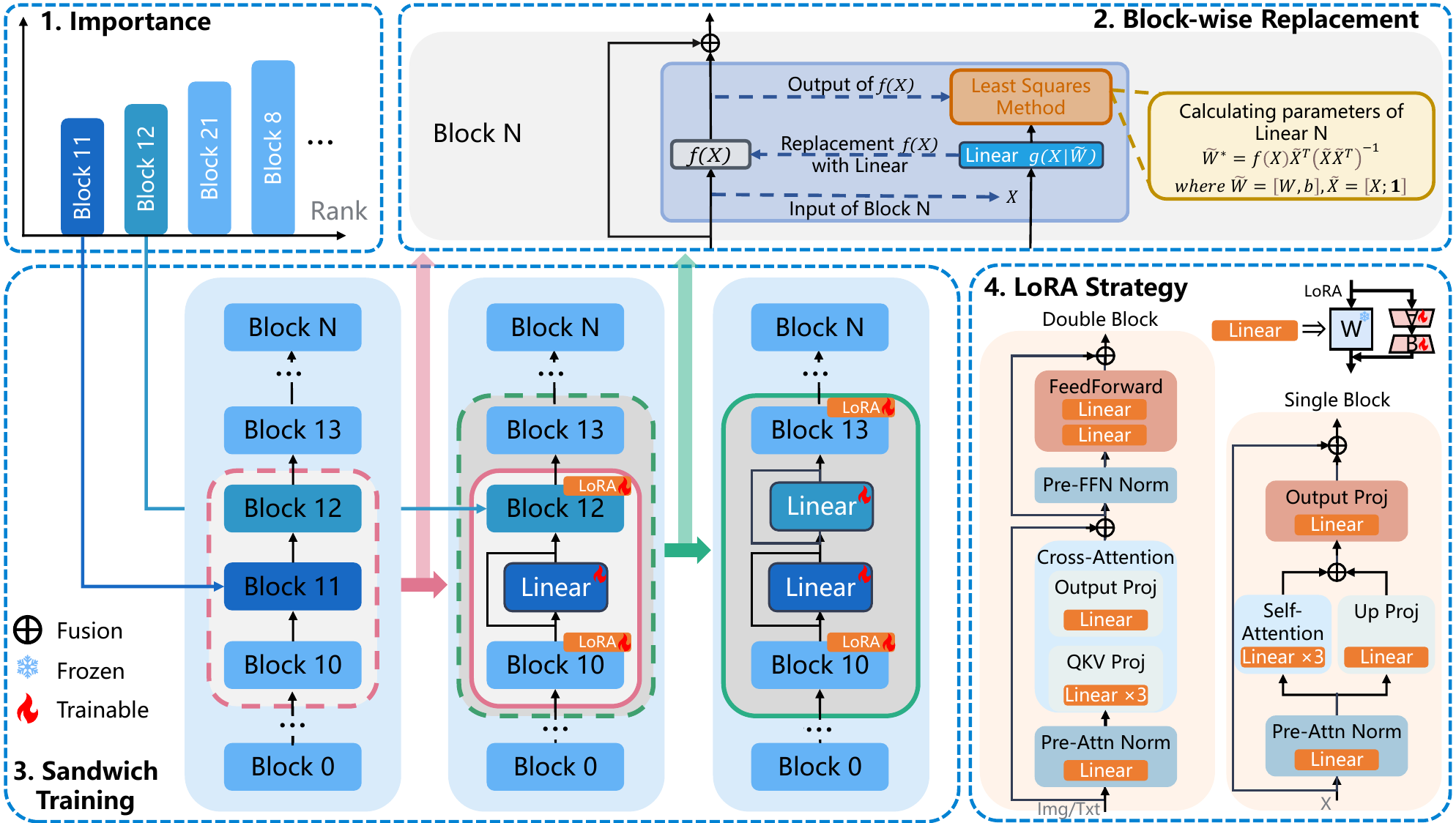}
  \caption{Overview of the FastFLUX Framework. \textbf{(1) Importance Estimation}: Each block is ranked by estimated importance to determine the pruning order.
 \textbf{(2) Block-wise Replacement with Linear Layers (BRLL)}: Selected blocks have their residual branches replaced with lightweight linear layers, whose parameters are initialized using a closed-form least-squares fit to the original block outputs.
  \textbf{(3) Sandwich Training (ST)}: For non-overlapping cases, we train the replaced block along with its immediate neighbors enhanced by LoRA. For overlapping cases, we search upward or downward to locate the nearest unpruned blocks, apply LoRA to those, and jointly train all intermediate replaced blocks. Input-output data for each training step are re-collected by loading all previously trained sandwich structures. \textbf{(4) LoRA Strategy}: LoRA modules are selectively inserted into either single or double blocks during ST to provide local supervision while keeping the rest of the model frozen. This framework enables efficient and structured pruning with minimal performance degradation.}
  \label{main_pic}
\end{figure}
\section{FastFLUX: Block-wise Efficient Pruning with Sandwich Training}
FastFLUX is a lightweight and effective framework for pruning large-scale diffusion transformers, improving inference efficiency while preserving high-quality generation.
We begin by introducing \textbf{Block-wise Replacement with Linear Layers (BRLL)}, which substitutes the computationally heavy residual branches in ResBlocks with linear approximations while keeping the original shortcut pathways intact to maintain stable information flow.
We then describe \textbf{Sandwich Training (ST)}, a localized fine-tuning strategy that applies LoRA-based supervision to the neighboring unpruned layers of each replaced block, providing targeted adaptation with minimal training overhead.
An overview of the full pipeline, together with implementation details, is provided in Figure~\ref{main_pic} and Algorithm~\ref{alg:detail}.

\subsection{Block-wise Replacement with Linear Layer}
\label{BRLL}

\begin{figure}[htbp]
  \centering
  \includegraphics[width=1\linewidth]{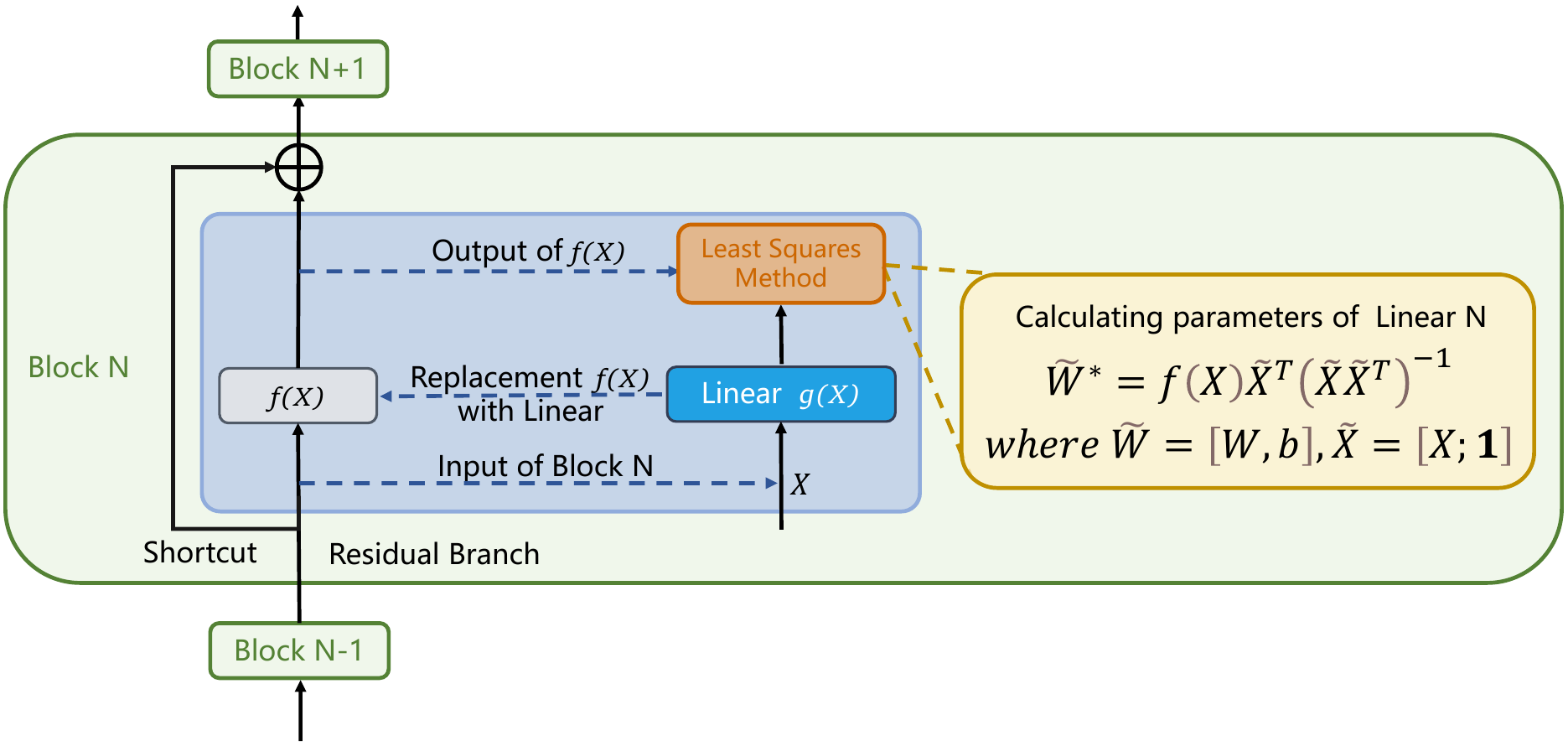}
  \caption{Illustration of the Block-wise Replacement with Linear Layers (BRLL) method.
  Each ResBlock consists of a residual branch and a shortcut branch. The residual branch is replaced with a linear layer, whose parameters are computed using input-output features collected during the forward pass. The shortcut connection is preserved to maintain stable information flow.}
  \label{fig_BRLL}
\end{figure}

To effectively compress ultra-large diffusion transformers such as FLUX, we propose a structural pruning strategy termed BRLL, which replaces the computationally expensive residual pathways of each block with lightweight linear modules. 
We observe that FLUX employs deep stacks of transformer-style ResBlocks, each of which consists of an expressive residual branch and a shortcut connection.
Our profiling reveals that the residual branches dominate both parameter count and FLOPs, making them the primary bottleneck for inference.
Motivated by this observation, BRLL substitutes only the residual branch with a lightweight linear approximation while preserving the original shortcut pathway.
Unlike replacing the entire block, which results in severe degradation due to the loss of the identity pathway, selective replacement preserves stable information flow and significantly reduces computational cost.
This distinguishes our approach from prior pruning methods that often disregard the functional asymmetry within ResBlocks. An overview of BRLL is provided in Figure~\ref{fig_BRLL}.

Formally, let \textit{X} denote the input to a ResBlock, $f(X)$ the output of its residual branch, and $Y = f(X) + X$ the original output of the ResBlock. 
To reduce computation while preserving the residual structure, we replace the residual branch with a lightweight linear layer $g(X) = WX + b$, resulting in the modified output $\hat{Y} = g(X) + X$.
To preserve the representational behavior of the original block, we minimize the approximation error between the original residual output and its linear approximation. This leads to the following objective: $\min_{W, b} || f(X) - (WX + b) ||^2$. 
The optimal parameters $W^*$ can be obtained using the least squares method based on the collected input-output pairs. By defining the augmented input $\tilde{X} = [X; \mathbf{1}]$ and the combined parameters $\tilde{W} = [W, b]$, the closed-form solution is:
\begin{equation}
\tilde{W}^* = f(X)\tilde{X}^T(\tilde{X}\tilde{X}^T)^{-1}.
\label{eq1}
\end{equation}
This procedure yields an efficient closed-form initialization of the linear layer's parameters, which can be further optimized during the subsequent Sandwich Training phase.

\begin{algorithm}[t]
\caption{Block-wise Replacement with Linear Layer and Sandwich Training.}
\label{alg:detail}
\SetKwInOut{Require}{Require}
\SetKwInOut{Output}{Output}

\Require{Pretrained FLUX model, replacement ratio $r \in [0, 1]$, the number of blocks in the model $N$, the weight of FID in computing the importance score $\alpha$, the weight of CLIP score in computing the importance score $\beta$.}
\Output{Pruned model with replaced linear blocks}

\BlankLine
\For{each block $B_i$}{
    Temporarily remove $B_i$ and conduct forward propagation on benchmark datasets; \\
    Compute FID and CLIP score; \\
    Compute the importance score $s_i = \alpha {\rm FID} + \beta {\rm CLIP}$;\\
}
Sort the importance scores $\{s_i\}_{i=1}^N$ in the ascending order;

Select the top-$r$ least important blocks $\{B_k\}_{k=1}^{rN}$ according to their importance scores\;

    \For{each $B_k \in \{B_k\}_{k=1}^{rN}$}{
        // Block-wise Replacement with Linear Layer  \\
        Collect the input and output of the residual branch in $B_k$; \\

        Compute the weight of the linear layer that is used to replace the residual branch: \\
        ~~~~~~~~~~~~~~~~~~~~$\tilde{W}^* = f(X)\tilde{X}^T(\tilde{X}\tilde{X}^T)^{-1}, \tilde{X} = [X; \mathbf{1}], \tilde{W} = [W, b]$ \\
        // Sandwich Training \\
        Take $B_k$ as the center and combine the ones before and after it to construct a \textbf{sandwich block}; \\

        Collect the input and output of the sandwich block for finetuning; \\

        Replace the residual branch of $B_k$ with the linear layer using the weights $W*$; \\
        
        Add LoRA into the first and last blocks inside the sandwich block; \\

        Finetune the sandwich by minimizing the difference between the original and pruned one. \\
        
    }

\end{algorithm}

\subsection{Sandwich Training}
\label{Sandwich Training}
To mitigate performance degradation after block replacement, we introduce \textbf{Sandwich Training (ST)}, a localized fine-tuning strategy that adapts each replaced block using contextual supervision from its nearest unpruned neighbors.
We observe that different blocks contribute unequally to the model’s performance, and prioritizing the pruning of less important blocks leads to more stable and effective compression. Therefore, we begin this section by introducing a method for \textbf{Estimating Block Importance}. Based on the computed importance scores, block pruning may lead to structural overlap between selected regions. To address this, our ST strategy includes dedicated mechanisms for \textbf{Training Sandwich Blocks without Overlap} and \textbf{with Overlap}, respectively. Finally, in \textbf{Parameter Reuse across Pruning Ratios}, we present an empirical finding that transferring trained parameters from high-pruned models to lower-pruned configurations can further improve performance without additional training.

\textbf{Block Importance Estimation}.
We observe that different blocks contribute unequally to the model’s generative performance (see supplementary materials for more results).
Based on this observation, we adopt an importance-aware pruning strategy that prioritizes the removal of less critical blocks.
To quantitatively assess the importance of each block, we combine two commonly used metrics in text-to-image evaluation: FID~\cite{heusel2018ganstrainedtimescaleupdate} and CLIP Score~\cite{radford2021learningtransferablevisualmodels}.
Specifically, for each ResBlock, we temporarily remove it from the model and generate a set of images. We then compute the FID and CLIP scores for each generated set and normalize both scores independently across all blocks.
Since a lower FID indicates better quality, we transform it by taking $1-FID_{norm}$ so that higher values are consistently better across both metrics.
The importance score $s_i$ for the i-th block is defined as:
\begin{equation}
    s_i = -(\alpha (1-{\rm FID_{norm}^{(i)}}) + \beta {\rm CLIP_{norm}^{(i)}}),
\end{equation}
where \(\alpha\) and \(\beta\) are hyperparameters that balance the contribution of each metric. 
In our experiments, we set $\alpha\ = \beta\ = 0.5$. 
A lower score $s_i$ implies that removing the block has a smaller impact on the overall generation quality, indicating that the block is less important and should be pruned earlier in the compression process. We therefore sort all blocks in ascending order of their importance scores and 
perform block replacement accordingly. 
Although Stable Flow~\cite{StableFlow} also employs layer-wise importance estimation, it relies solely on similarity-based metrics. 
In contrast, we incorporate both similarity and image quality assessments, enabling a more comprehensive evaluation of each block's contribution to generation performance.

\begin{figure}[t]
  \centering
  \includegraphics[width=1\linewidth]{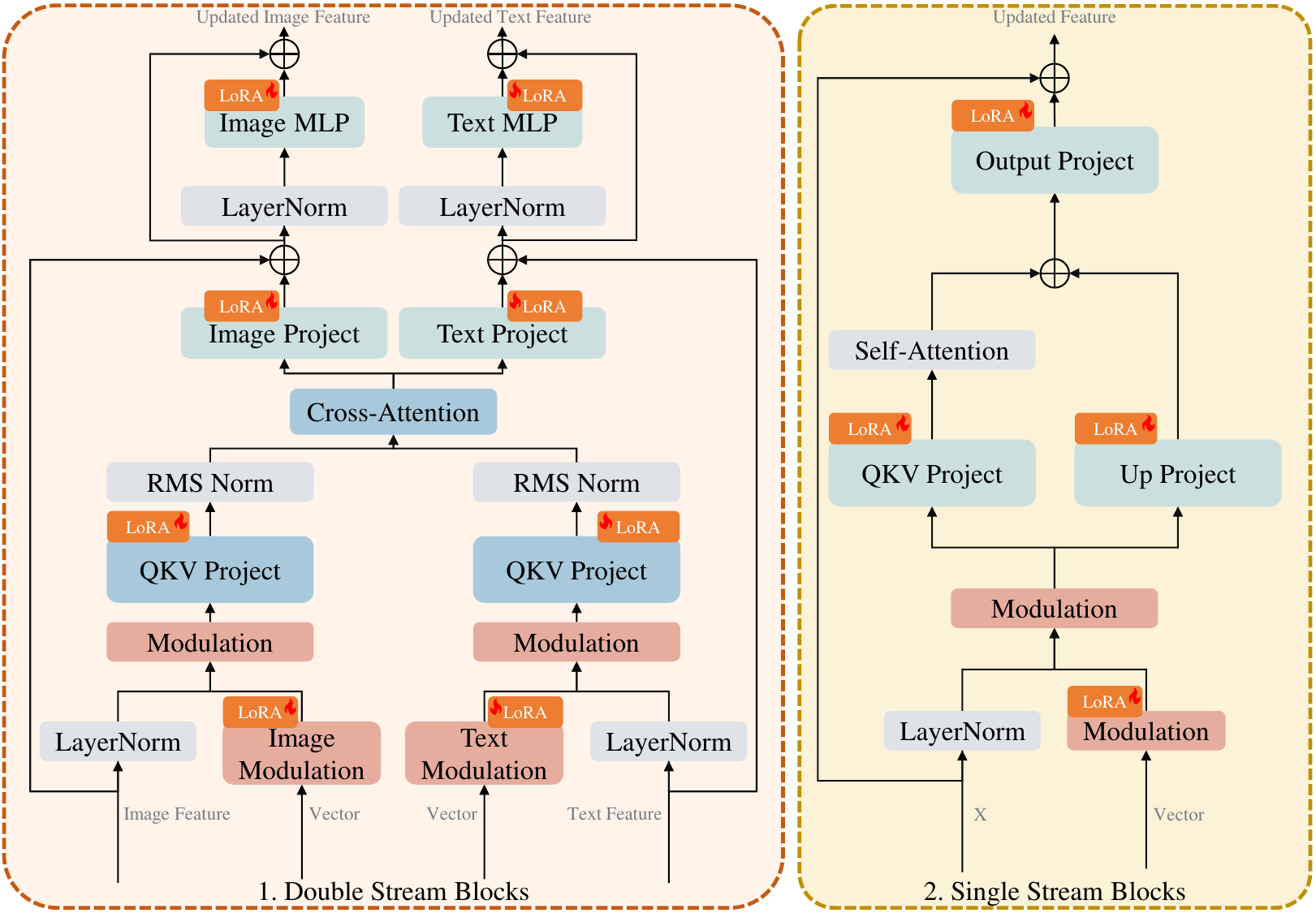}
    \caption{Illustration of the LoRA insertion strategy used in our Sandwich Training (ST). For each block selected for replacement with a linear layer, we identify its closest unpruned neighboring blocks before and after. LoRA modules are inserted into these adjacent blocks to provide localized supervision, enabling efficient and targeted fine-tuning of the replaced block.}
  \label{block_apply_lora}
\end{figure}

\textbf{Training Sandwich Blocks without Overlap}. 
When a ResBlock’s residual branch is replaced with a single linear layer, the number of trainable parameters becomes significantly limited. 
To better align the pruned model with the original network and mitigate accuracy degradation caused by pruning, we introduce the ST strategy.
The key idea is to leverage the neighboring blocks to enhance the training of the locally replaced block.
Specifically, when the block to be replaced is $\mathbf{B}_{i}$, we first collect training data by recording the input to $\mathbf{B}_{i-1}$ and the output from $\mathbf{B}_{i+1}$ during a forward pass. 
Then, we isolate three blocks: $\mathbf{B}_{i-1}$, $\mathbf{B}_{i}$, $\mathbf{B}_{i+1}$. We replace the residual branch of $\mathbf{B}_{i}$ with a linear layer, apply LoRA to $\mathbf{B}_{i-1}$ and $\mathbf{B}_{i+1}$, and allow the full parameters of the linear layer in $\mathbf{B}_{i}$ to be trainable.
The structure of the LoRA injection is illustrated in Figure~\ref{block_apply_lora}.
The collected input-output pair forms a localized training dataset. 
This sandwich-like configuration, where the replaced linear layer is flanked by two LoRA-enhanced blocks, is shown in the Sandwich Training part in Figure~\ref{main_pic}. 
Each linear layer is initialized using the solution described in Section~\ref{BRLL}.
Before pruning a new block, we reload all previously trained sandwich structures to collect updated input-output pairs from the current model. 
This ensures that each new sandwich block is trained with data that reflects the latest model state, promoting consistency and stability during progressive pruning.
This formulation above applies to the non-overlapping case, where both adjacent blocks have not yet been replaced, i.e., $\mathbf{B}_{i-1} \notin \mathcal{R}$ and $\mathbf{B}_{i+1} \notin \mathcal{R}$, with $\mathcal{R}$ denoting the set of previously replaced blocks. This is shown in the middle column of the Sandwich Training part in Figure \ref{main_pic}.

\textbf{Training Sandwich Blocks with Overlap}. 
When at least one adjacent block of the current block $\mathbf{B}_{i}$ has already been replaced with a linear layer, that is, $\mathbf{B}_{i-1} \in \mathcal{R}$ or $\mathbf{B}_{i+1} \in \mathcal{R}$, the resulting sandwich structure overlaps with previously replaced blocks.
In such cases, we construct an extended sandwich structure by identifying the nearest unpruned blocks above and below $\mathbf{B}_{i}$ and applying LoRA to them.
Formally, we define:

\begin{equation}
u=\max \left\{j<i \mid \mathbf{B}_j \notin \mathcal{R}\right\}, \quad d=\min \left\{j>i \mid \mathbf{B}_j \notin \mathcal{R}\right\}
\end{equation}

where $u$ and $d$ represent the indices of the nearest unpruned blocks before and after $\mathbf{B}_{i}$, respectively. 
$j \in \mathcal{R}$ refers to a previously replaced block.
LoRA modules are applied to $\mathbf{B}_{u}$ and $\mathbf{B}_{d}$, while the linear layer in $\mathbf{B}_{i}$, together with all previously replaced linear layers between $\mathbf{B}_{u}$ and $\mathbf{B}_{d}$, are jointly trained using input-output data collected from the model with all prior sandwich structures loaded.
The overlap-aware configuration is illustrated in the right column of the ST section in Figure~\ref{main_pic}.

\textbf{Inference Reuse from High-Pruned Models}.
Our training strategy operates in a block-wise manner, where each replacement is optimized locally rather than through end-to-end fine-tuning. 
This design significantly reduces computational cost and memory consumption. 
As described earlier, before replacing a new block, we load all previously trained sandwich structures, including both linear layers and LoRA modules, and perform a forward pass to collect the input-output pairs required for training the current block. 
This ensures that each stage of training is based on the current model state, maintaining pruning stability and minimizing cumulative performance degradation.
At inference time, we assemble the final pruned model by incorporating all trained sandwich structures. 
An interesting phenomenon emerges when components trained under a higher pruning ratio model are reused in a lower pruning ratio configuration. 
Specifically, after training a model with a pruning ratio of $a\%$, we continue training it to a higher pruning ratio $b\%$, with $b>a$.
After the higher pruning stage is completed, we extract the corresponding trained components, including linear layers and LoRA modules, and insert them into the $a\%$ model without any additional fine-tuning. 
This reuse consistently leads to improved inference performance compared to using only the components trained at the $a\%$ stage.
We hypothesize that this improvement arises from block-level overlap across pruning stages. 
In higher pruning configurations, overlapping blocks participate in multiple sandwich training setups and are therefore updated more frequently with diverse input distributions. 
This repeated exposure enhances their generalization ability and alignment with the global structure of the model. 
When reused in the lower pruning configuration, these components offer improved representational capacity and lead to better generation quality without additional training.

\begin{table}[t]
\centering
\caption{Quantitative comparison on the HPS v2 benchmark. We group models by similar inference latency to fairly compare the effectiveness of FastFLUX and L1-norm pruning at different compression ratios. Results show that FastFLUX achieves better or comparable performance across all categories, and even outperforms the original uncompressed FLUX.1-dev model when pruned by 10 percent. \underline{Underlined} values indicate the best result in each category across all models, while \textbf{bold} values highlight the best result within each latency group.}
\resizebox{\linewidth}{!}{
\begin{tabular}{lccccc|c}
\toprule
\textbf{Model} & \textbf{Animation} & \textbf{Concept-art} & \textbf{Painting} & \textbf{Photo} & \textbf{Averaged} & \textbf{Latency (ms)$\downarrow$}\\
\midrule

DALL·E mini\cite{ramesh2021zero} & 26.10   & 25.56  & 25.56  & 26.12  & 25.83 & - \\

SD v1.4\cite{rombach2022highresolutionimagesynthesislatent} & 27.26  & 26.61  & 26.66  & 27.27  & 26.95  & -  \\
SD v2.0\cite{rombach2022highresolutionimagesynthesislatent} & 27.48  & 26.89  & 26.86  & 27.46  & 27.17& -  \\
SD v3.0\cite{esser2024scaling} & 28.58  & 27.87  & 28.15  & 27.72  & 28.08& -  \\
FLUX.1-dev  & 28.94 & 27.90 & 28.16 & 28.28 & 28.32
 & 98.26 \\
\bottomrule
FLUX-Pruning-12\%
     & 28.83 & 27.75 & 28.03 & 28.21 & 28.20 & 89.98 \\
FastFLUX-10\%  & \underline{\textbf{29.04}}  & \underline{\textbf{28.24}} & \underline{\textbf{28.49}}  & \underline{\textbf{28.57}} & \underline{\textbf{28.58}} & 90.78\\
\bottomrule
FLUX-Pruning-21\%     & 28.41 & 27.39& 27.67 & 27.91 & 27.84 & 85.66   \\
FastFLUX-15\%  & \textbf{28.58}  & \textbf{27.87} & \textbf{28.13}  & \textbf{28.19} & \textbf{28.19} & {85.39}\\
\bottomrule
FLUX-Pruning-27\%     & 27.66 & 26.67  & 27.03 & 27.31 & 27.17  & 81.98 \\
FastFLUX-20\%  & \textbf{28.06} & \textbf{27.48} & \textbf{27.72} & \textbf{27.77} & \textbf{27.76} & 80.83\\
\bottomrule
\end{tabular}}
\label{Quantitative:HPS}
\end{table}
\begin{table}[t]
\centering
\caption{Quantitative results on the GENEVAL benchmark. We compare two compression methods, FastFLUX and L1-norm pruning, under different pruning ratios. Models with similar inference latency are grouped together for fair comparison. FastFLUX consistently achieves better performance across most categories. \underline{Underlined} values indicate the best result in each category across all models, while \textbf{bold} values highlight the best result within each latency group.}
\resizebox{\linewidth}{!}{
\begin{tabular}{lccccccc|cc}
\toprule
\textbf{Model} & \textbf{Single} & \textbf{Two} & \textbf{Counting} & \textbf{Colors} & \textbf{Position} & \textbf{Attribute} & \textbf{Overall}  & \textbf{Latency (ms)$\downarrow$} \\
& \textbf{object} & \textbf{object} & & & & \textbf{binding} &  &\\
\midrule
DALL·E mini\cite{ramesh2021zero}      & 0.73 & 0.11 & 0.12 & 0.37 & 0.02 & 0.01 & 0.23  &-\\
SD v1.4\cite{rombach2022highresolutionimagesynthesislatent}         & 0.98  & 0.36 &  0.35 & 0.73 & 0.01 & 0.07 & 0.42 &- \\
SD v2.0\cite{rombach2022highresolutionimagesynthesislatent}         &0.98 & 0.50 & 0.48 & \underline{0.86} & 0.06 & 0.15 & 0.51 &- \\
SD v3.0\cite{esser2024scaling}          &0.98 & 0.75 & 0.51 & 0.84 & \underline{0.22} & \underline{0.52} & 0.63  &-\\
FLUX.1-dev    & \underline{0.99} & \underline{0.78} & \underline{0.70} & 0.78 & 0.19 & 0.46 & \underline{0.64} & 98.28\\
\bottomrule
FLUX-Pruning-12\%     & 0.98 & \textbf{0.72} & 0.59 & \textbf{0.76} & 0.15  & 0.40 & 0.60  &90.00 \\
FastFLUX-10\%    & \textbf{0.99} & 0.69 & \textbf{0.63} & 0.74 & \textbf{0.15} & \textbf{0.42} & \textbf{0.60}  & 90.85\\
\bottomrule
FLUX-Pruning-21\%    & 0.98 & 0.61 & 0.54 & \textbf{0.76} & 0.13  & 0.29 & 0.55  & 85.22\\
FastFLUX-15\%   & \textbf{0.98} & \textbf{0.65} & \textbf{0.56} & 0.73 & \textbf{0.13} & \textbf{0.35} & \textbf{0.57}  & 85.67\\
\bottomrule
FLUX-Pruning-27\%    & 0.97 & 0.43 & 0.44 & \textbf{0.71} & 0.09  & 0.19 & 0.47  & 81.59\\
FastFLUX-20\%   & \textbf{0.98} & \textbf{0.59} & \textbf{0.48} & 0.68 & \textbf{0.12}  & \textbf{0.31} & \textbf{0.53}  & 80.78\\
\bottomrule
\end{tabular}
}
\label{Quantitative:geneval}
\end{table}

\section{Experiments}
In this section, we experimentally validate the effectiveness and rationality of our proposed method. 
We begin by introducing the experimental setup and implementation details. 
Then, we conduct quantitative, qualitative, and efficiency comparisons with existing T2I methods to assess overall performance. 
Furthermore, we present generalization experiments and ablation studies to provide deeper insights into the design behind FastFLUX.
Finally, we include several additional experiments to further strengthen the completeness and logical rigor of our methodology.
\begin{figure}[htbp]
  \centering
  \includegraphics[width=0.9\linewidth]{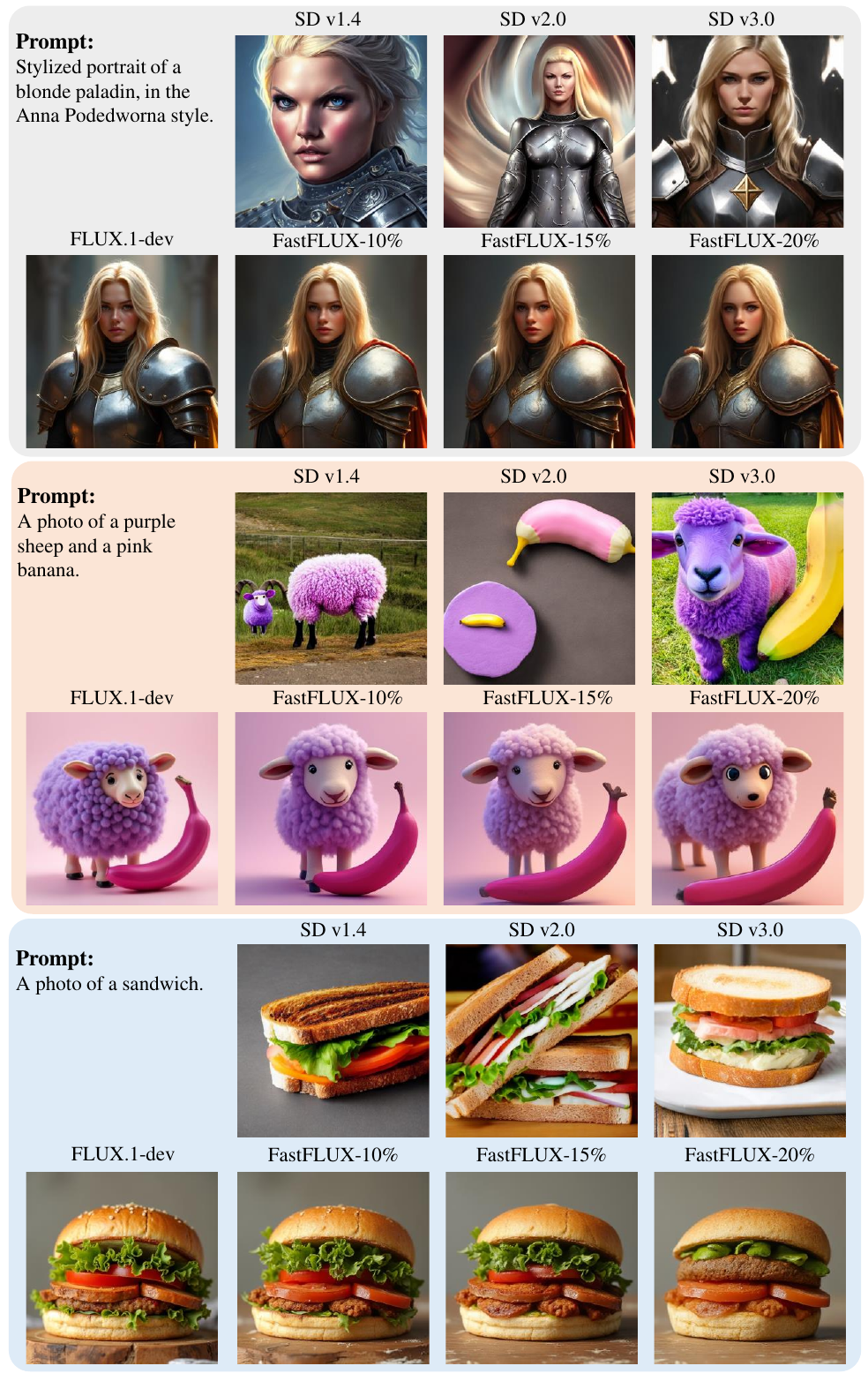}
  \caption{Qualitative comparison between FastFLUX and other mainstream T2I models under the same prompts. FastFLUX consistently generates images of higher visual fidelity. Even at different pruning ratios, the outputs of FastFLUX remain similar to those of the original FLUX.1-dev, while outperforming non-FLUX baselines in overall quality.}
  \label{Visual2}
\end{figure}
\subsection{Implementation Details}
Our base model is FLUX.1-dev, and all experiments are conducted using the fluxdev-controlnet-16k dataset~\cite{fluxdevcontrolnet16k}, which consists of approximately 16.1k images.
We sample 512 images to compute the parameters of the linear layers and another 512 images for training.
Since we only need to fit the parameters of a single linear layer and a few LoRA modules at a time, only a small number of samples are required for training. We set the rank to 16 and the alpha to 8 of LoRA.
To provide a comprehensive and holistic assessment of our method, we utilize a range of text-to-image evaluation benchmarks, including HPS v2~\cite{wu2023humanpreferencescorev2} and GENEVAL~\cite{ghosh2023genevalobjectfocusedframeworkevaluating}.All experiments are implemented in PyTorch and executed on two NVIDIA RTX 3090 GPUs using a block-swap mechanism for memory-efficient training.

\subsection{Performance Comparison}

\begin{figure}[t]
  \centering
  \includegraphics[width=1\linewidth]{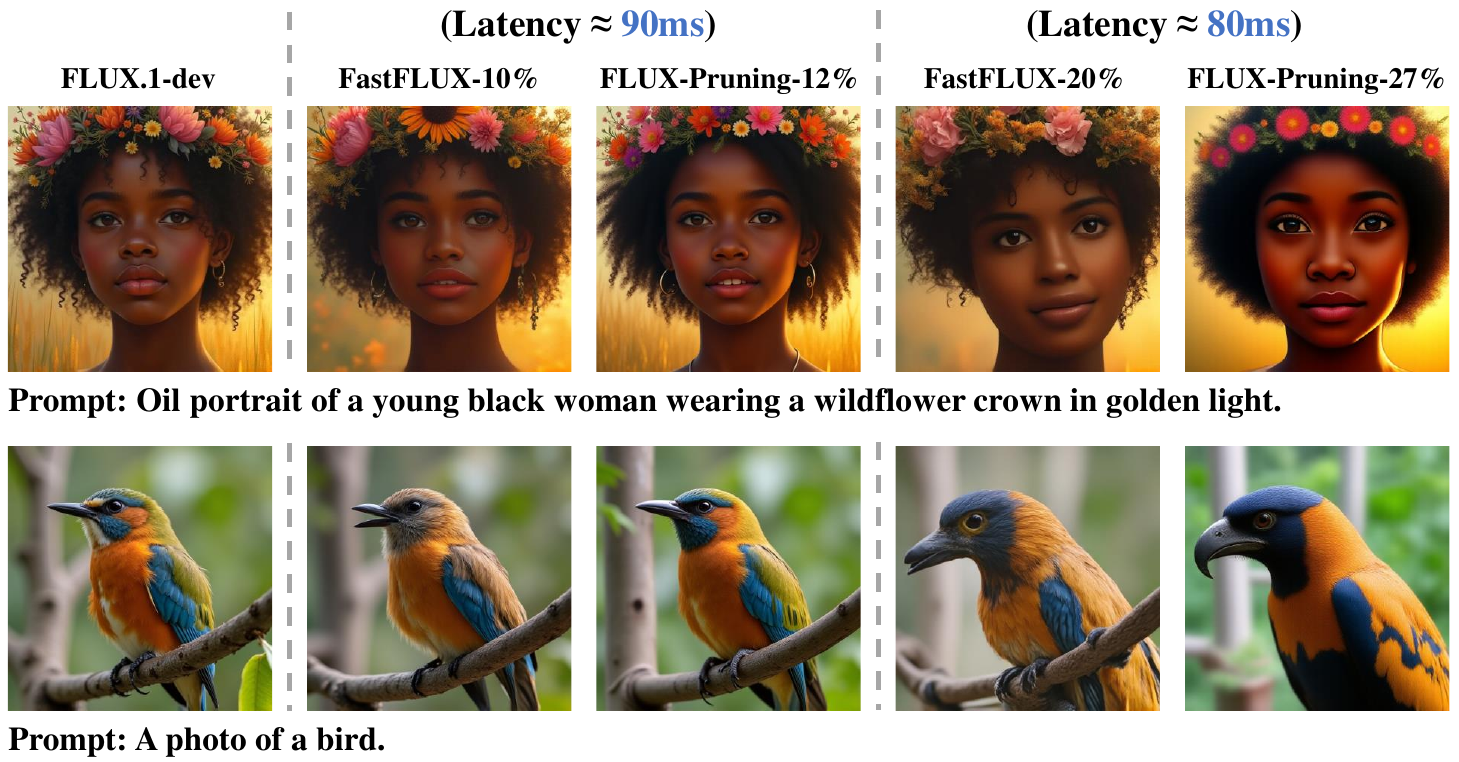}
  \caption{Qualitative comparison between FastFLUX and L1-norm pruned models under similar latency settings. Within each latency group, FastFLUX achieves better alignment with the original FLUX.1-dev output, preserving more texture and visual detail. 
  In contrast, L1-norm pruned models tend to lose fine-grained information, especially under higher pruning ratios. 
  }
  \label{Visual1}
\end{figure}

\noindent\textbf{Quantitative Results on Multiple Benchmarks}.
We evaluate the T2I generation performance of our model on two widely used benchmarks: 
HPS v2 (for assessing alignment with complex human preferences) and GENEVAL (for evaluating short-prompt generation). The results are shown in Table\ref{Quantitative:HPS} and Table\ref{Quantitative:geneval}, respectively.
We compare FastFLUX with the original FLUX.1-dev, L1-norm pruned variants of FLUX, and several state-of-the-art T2I models\cite{ramesh2021zero,rombach2022highresolutionimagesynthesislatent,rombach2022highresolutionimagesynthesislatent,esser2024scaling}. 
Since both FastFLUX and the L1-norm method are model compression techniques, we group models by similar inference latency for fair comparison.
From the results, we observe that FastFLUX, when pruned by 10 percent, not only preserves generation quality but also outperforms the original FLUX.1-dev on the HPS v2 benchmark. Specifically, it achieves an improvement of 0.26 in score and ranks first among all evaluated T2I models. In addition, FastFLUX reduces inference latency by 7.48ms compared to the unpruned baseline.
Within the same latency group, the L1-norm pruned model removes 12 percent of the parameters but performs 0.38 lower than FastFLUX on the HPS v2 metric. 
Moreover, L1-based pruning requires global fine-tuning to adapt to the irregular parameter structure, resulting in higher training costs. 
In contrast, FastFLUX requires only minimal localized adaptation.
These observations are consistent across other latency groups and evaluation settings. 
Overall, FastFLUX significantly improves inference efficiency while maintaining high-quality image generation.

\noindent\textbf{Qualitative Results on Generated Samples}.
For qualitative evaluation, we present a visual comparison of image generation quality across different models. As shown in Figure \ref{Visual2}, FastFLUX consistently produces images of higher quality compared to other mainstream text-to-image models. Among the FastFLUX variants, even under varying pruning ratios, the generated results remain visually similar to those from the original FLUX.1-dev model. Although higher pruning ratios may lead to the loss of fine details, the visual quality of FastFLUX remains superior to that of non-FLUX baselines.
In Figure\ref{Visual1}, we compare the visual outputs of FastFLUX and the models pruned via L1-norm. When inference latency is almost the same, FastFLUX is able to retain more texture and visual details, showing better alignment with the unpruned model. In contrast, the L1-norm pruned models tend to lose more fine-grained information, especially at higher pruning ratios. These results demonstrate that FastFLUX effectively balances inference efficiency with high-fidelity image generation.

\noindent\textbf{Efficiency Analysis}.
In Table\ref{Quantitative:HPS} and Table\ref{Quantitative:geneval}, we group models with similar inference latency together for fair comparison.
The quantitative results show that although latency is comparable, the L1-norm-based pruning approaches are consistently outperformed by our proposed FastFLUX in terms of both image generation quality and pruning effectiveness. Moreover, the L1-norm method requires full model fine-tuning after pruning, which leads to significantly higher training cost. In contrast, FastFLUX achieves a better trade-off between efficiency and quality, making it a more practical and scalable solution for model acceleration.

\noindent\textbf{Generalization to Other Architectures}.
While FastFLUX is primarily evaluated on the FLUX model with a DiT-style backbone, we further assess its generalization capability on Stable Diffusion v3.0~\cite{esser2024scaling}, a model with a different architectural design.
Specifically, we apply both BRLL and ST to SD v3.0, which differs in depth, attention layout, and residual structure.
As shown in Table~\ref{Generality}, the FastSD v3.0-10\% variant achieves a higher CLIP Score compared to the unpruned SD v3.0 baseline, with only a moderate drop in HPS v2. These results confirm that BRLL and ST generalize well beyond DiT-style transformers and can benefit other class of diffusion-based generative models.
\begin{table}[t]
\centering
\caption{Generalization of our method to the SD v3.0 model. The improved CLIP score on FastSD demonstrates that our method remains effective when applied to other text-to-image models beyond the FLUX framework.}
\vspace{4pt}
\resizebox{0.5\linewidth}{!}{
\begin{tabular}{lcc}
\toprule
\textbf{Model} & \textbf{HPS v2} & \textbf{CLIP Score}  \\
\midrule
FLUX.1-dev   & 28.32 & 31.35  \\
FastFLUX-10\%   & \textbf{28.58} & 30.98   \\
SD v3.0\cite{esser2024scaling}   & 28.08 & 31.79 \\
FastSD v3.0-10\%  & 27.76 & \textbf{31.82}  \\

\bottomrule
\end{tabular}}
\label{Generality}
\end{table}

\begin{figure}[hbpt]
  \centering
  \includegraphics[width=1\linewidth]{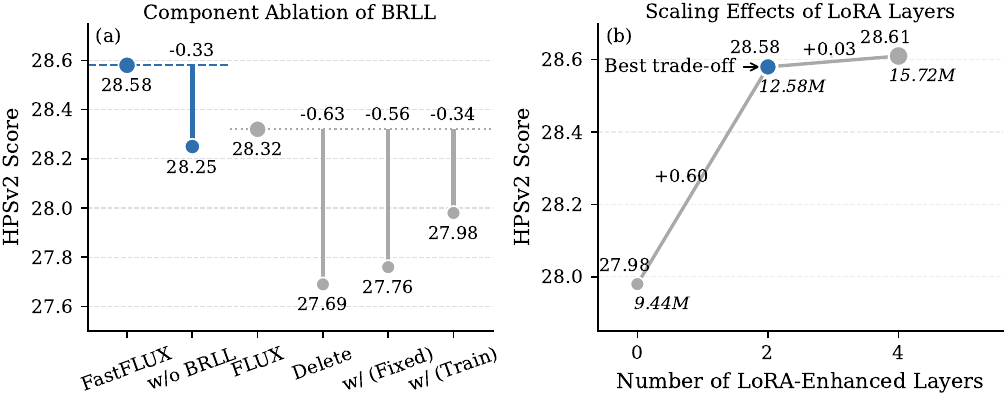}
    \caption{(a) Component ablation of BRLL on FastFLUX and FLUX. w/o: FastFLUX without BRLL. Del: FLUX with block deletion. w/ F: FLUX with deleted blocks and fixed BRLL. w/ T: FLUX with deleted blocks and trained BRLL.
(b) Effects of the number of LoRA-enhanced layers in ST. The x-axis indicates the total count of LoRA-enhanced layers. Performance improves substantially from 0 to 2 LoRA layers and saturates at 4, indicating a 2-layer structure provides a trade-off between performance and efficiency.}
  \label{Lora sandwich layers}
\end{figure}

\begin{table}[htbp]
\centering
\caption{Ablation study on the effectiveness of the Sandwich Training (ST) strategy. Without ST, the model applies BRLL to replace residual branches with linear layers using calculated parameters, but without further adaptation. As shown, omitting ST leads to a notable drop in HPS v2 performance.}
\vspace{4pt}
\resizebox{0.5\linewidth}{!}{
\begin{tabular}{lcccc}
\toprule
\textbf{Method} & \textbf{HPS v2} & \textbf{CLIP Score}  & \textbf{Latency (ms)$\downarrow$} \\
\midrule
FLUX.1-dev  & 28.32 & \textbf{31.35}  & 98.26 \\
FastFLUX-10\%  &  \textbf{28.58} & 30.98  & \textbf{90.78} \\
w/o ST  & 27.99 & 30.91  & 90.83\\
\bottomrule
\end{tabular}}
\label{Effic:sandwich}
\end{table}
\vspace{4pt}

\begin{table}[t]
\centering
    \caption{Comparison of different block replacement orders. Our importance-guided strategy achieves the best performance, while the Strat2End strategy performs worst, likely due to the generally higher importance of double blocks compared to single blocks. See Figure~\ref{G1} for details.}
\vspace{4pt}
\resizebox{0.5\linewidth}{!}{
\begin{tabular}{lccc}
\toprule
  & \textbf{FastFLUX} & \textbf{Start2End} & \textbf{End2Strat}  \\
\midrule
HPSv2  & \textbf{28.58} & 24.63 & 28.15  \\
\bottomrule
\end{tabular}}
  \label{Effic:train in importance}
\end{table}

\begin{table}[t]
\centering
\caption{Performance under different block replacement ratios. As the ratio increased, the reduction in computational complexity and inference time became more significant, while the decrease in HPS v2 was relatively moderate.}
\vspace{4pt}
\resizebox{0.8\linewidth}{!}{
\begin{tabular}{lcccc}
\toprule
\textbf{Ratio} & \textbf{Drop \#Block} & \textbf{FLOPs (T)} & \textbf{Inference Time(ms)} & \textbf{HPSv2} \\
\midrule
0 (baseline)  & 0 & 38.19 & 98.26 & 28.32  \\
5\%  & 3 & 36.79 \textbf{(96.3\%)} & 93.24 & 28.23 \\
10\%  & 5 & 35.87 \textbf{(93.9\%)} & 90.78 & \textbf{28.58} \\
15\%  & 8 & 33.39 \textbf{(87.4\%)} & 85.39 & 28.19 \\
20\%  & 11 & 32.46 \textbf{(85.0\%)} & 80.83 & 27.76 \\
25\%  & 14 & 31.07 \textbf{(81.3\%)} & 76.85 & 27.10  \\
30\%  & 16 & 29.37 \textbf{(76.9\%)} & 72.30 & 26.70  \\
\bottomrule
\end{tabular}}
\label{Avlation: drop block ratio}
\end{table}

\subsection{Ablation Studies}
\label{Ablation}

In this section, we conduct a series of ablation studies to evaluate the key components of our method. 
These include the impact of block-wise replacement with linear layers, the effect of the proposed sandwich training strategy, the influence of the number of layers used to construct the sandwich block, the impact of different pruning orders, and the performance under varying pruning ratios.

\noindent\textbf{Impact of Block-wise Replacement with Linear Layer}. 
Figure~\ref{Lora sandwich layers}(a) analyzes the effect of BRLL under different ablation settings. Removing BRLL from FastFLUX causes a performance drop, with HPSv2 decreasing from 28.58 to 28.25, indicating that BRLL is essential for preserving model capacity. In contrast, directly deleting selected blocks in FLUX results in a more severe degradation. Introducing BRLL partially alleviates this issue: fixed BRLL parameters yield a moderate performance recovery, while training BRLL further improves performance. These results demonstrate that BRLL is effective and plays a crucial role in recovering model performance under structural modification.

\noindent\textbf{Impact of Sandwich Training}.
We evaluate the effectiveness of the proposed ST strategy by comparing it with a variant that only applies the initialization described in Section~\ref{BRLL}, without any further fine-tuning.
As shown in Table\ref{Effic:sandwich}, removing ST leads to a notable drop in HPS v2 score from 28.58 to 27.99, resulting in a 0.59 degradation. 
This demonstrates that solely relying on the initial linear approximation is insufficient to fully preserve generative quality.
By allowing neighboring blocks with linear replacements to participate in the training process through LoRA modules, the ST strategy introduces localized supervision that helps compensate for performance loss caused by structural pruning.
These results confirm that ST is both effective and essential for maintaining model quality under architectural modifications.

\noindent\textbf{Number of layers in Sandwich Block}.
In FastFLUX, Sandwich Training applies LoRA modules to adjacent blocks around the replaced linear layer to provide localized supervision. 
We investigate how the number of surrounding layers involved in ST affects model performance.
Specifically, a setting of 0 layers corresponds to applying no LoRA modules.
A 2-layer configuration includes LoRA on $\mathbf{B}_{i-1}$ and $\mathbf{B}_{i+1}$, while a 4-layer setting extends this to $\mathbf{B}_{i-2}$, $\mathbf{B}_{i-1}$, $\mathbf{B}_{i+1}$, and $\mathbf{B}_{i+2}$.
As shown in Figure~\ref{Lora sandwich layers}(b), moving from 0 to 2 layers yields a substantial improvement in HPS v2 score by 0.60, demonstrating the benefit of local supervision from adjacent blocks. However, increasing from 2 to 4 layers results in only a marginal gain 0.03, while introducing a 25\% increase in trainable parameters. This diminishing return, coupled with increased computational cost, makes the 2-layer configuration a favorable trade-off between performance and efficiency.

\begin{figure}[hbpt]
  \centering
  \includegraphics[width=1\linewidth]{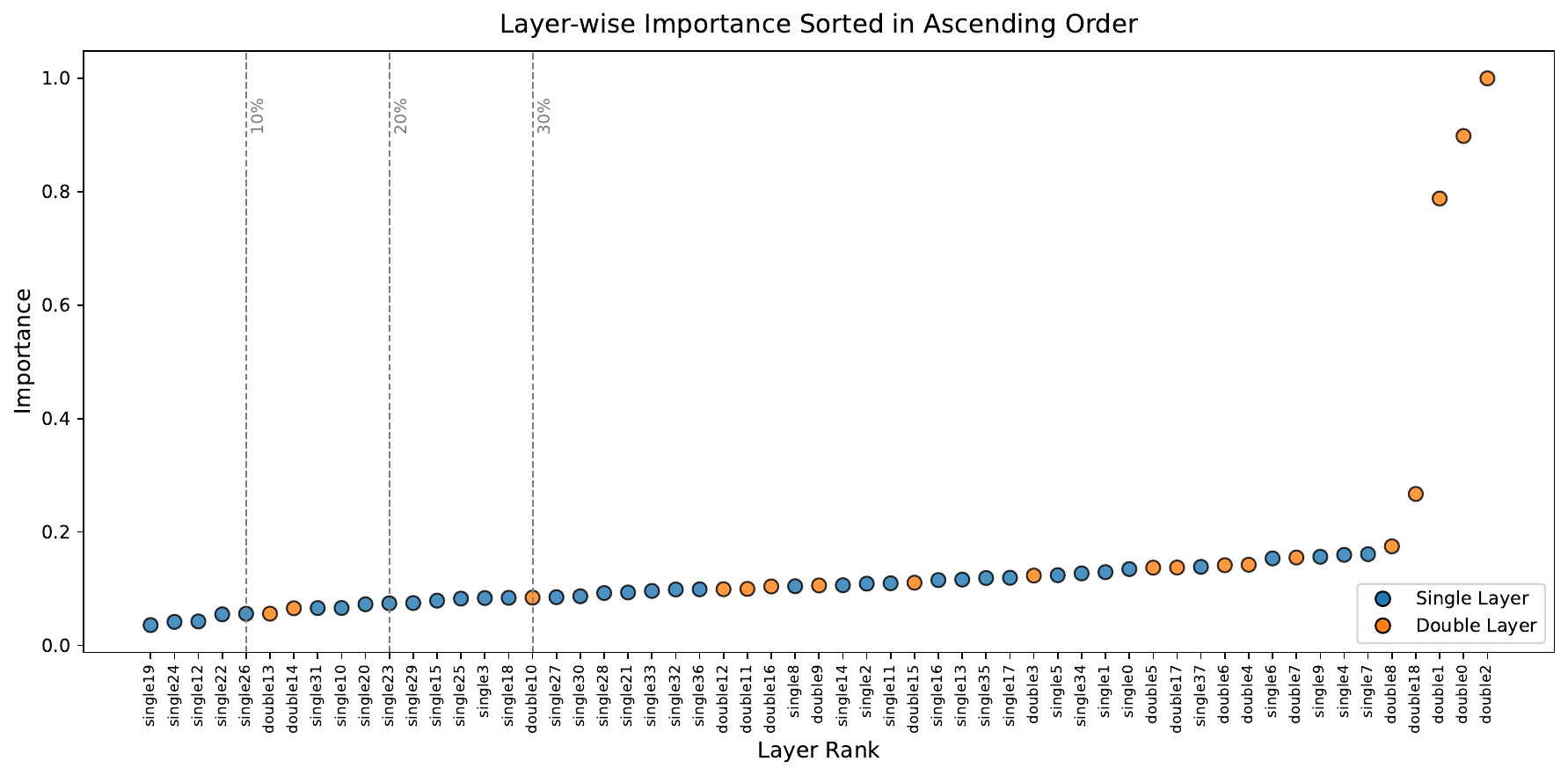}
    \caption{Importance scores of all layers, sorted in ascending order. Each dot represents a layer, with yellow indicating double layers and blue indicating single layers. The dashed vertical lines at 10\%, 20\%, and 30\% mark the thresholds used for Block-wise Replacement with Linear Layers.}
  \label{G1}
\end{figure}

\begin{figure}[t]
  \centering
  \includegraphics[width=0.9\linewidth]{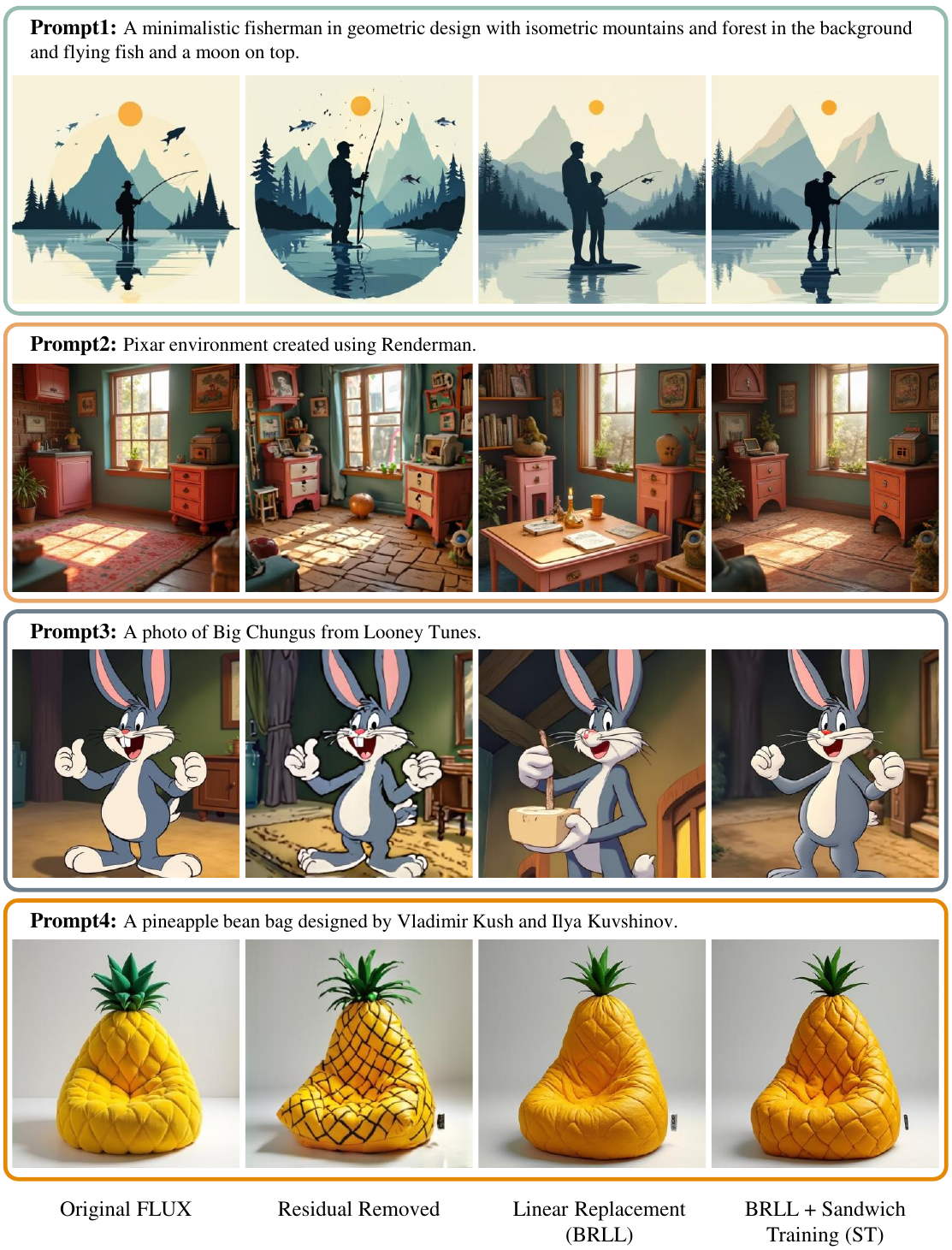}
    \caption{Visual comparison of image generations across four prompts (rows) and four methods (columns). From left to right: baseline output, result after direct layer deletion, result using naive linear replacement without our sandwich training strategy, and our proposed method. Our method best preserves the original visual characteristics of the baseline, demonstrating the effectiveness of both \textbf{BRLL} and \textbf{ST} in maintaining image fidelity.}
  \label{compare}
\end{figure}

\noindent\textbf{Impact of the Order of Blocks to be Pruned}.
A key design in FastFLUX is to prioritize pruning blocks with lower estimated importance, thereby minimizing performance degradation. 
To assess the impact of pruning order, we compare our importance-guided strategy with two sequential baselines: Start2End and End2Start. 
The former prunes blocks following the data flow direction (from shallow to deep), while the latter does so in the reverse direction.
As shown in Table~\ref{Effic:train in importance}, the importance-based strategy achieves the best HPS v2 score of 28.58, outperforming End2Start and Start2End by 0.43 and 3.95 points, respectively. 
These results demonstrate that the choice of pruning order significantly influences model performance.
We hypothesize that pruning in a Start2End fashion may remove low-level blocks that are critical for extracting foundational visual features. 
In contrast, End2Start pruning disrupts high-level semantics and weakens the gradient signals needed to update shallow layers effectively.
FastFLUX avoids these pitfalls by selecting blocks based on calculated importance scores rather than depth. 
This fine-grained, contribution-aware strategy enables a better balance between compression and generative quality.

\noindent\textbf{Performance with Different Pruning Ratios}. 
We apply FastFLUX to prune the FLUX model under various pruning ratios.
As shown in Table~\ref{Avlation: drop block ratio}, increasing the pruning ratio leads to a clear reduction in both computational cost and inference latency. 
For example, at a 30\% pruning ratio, FLOPs are reduced to 76.9\% of the original model and inference time drops by 25.96 ms, while the HPS v2 score decreases by only 1.62.
A particularly noteworthy observation is the asymmetry between performance degradation and efficiency gain. 
While inference time and FLOPs decrease significantly with higher pruning ratios, the HPS v2 score remains relatively stable up to a moderate pruning level (e.g., 10–15\%). This highlights the efficiency and robustness of the FastFLUX pruning strategy, which enables substantial acceleration with minimal quality loss.

\subsection{Additional Experiments}
\label{add_exp}
To further support the rationale and effectiveness of our method, we present two complementary experiments.
In Layer Importance Analysis, we quantitatively evaluate the contribution of each block to generation quality, providing the basis for our importance-aware pruning strategy.
In Visual Analysis of Multiple Replacement Strategies, we qualitatively compare different block replacement methods, demonstrating the visual advantages of combining BRLL with ST.

\begin{figure}[t]
  \centering
  \includegraphics[width=1\linewidth]{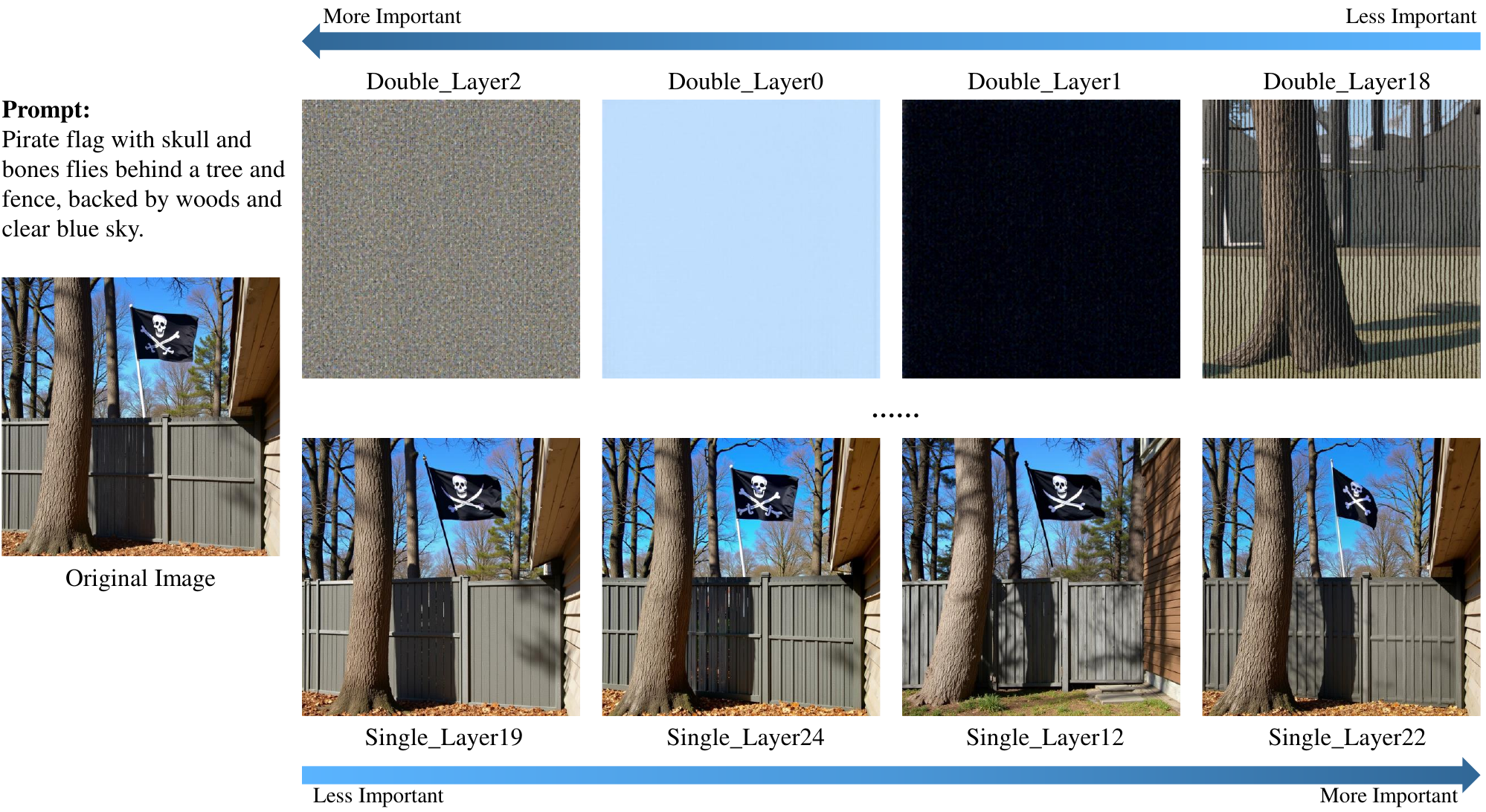}
    \caption{Visual impact of removing layers with different importance scores. The leftmost image shows the original output generated by the FLUX.1-dev model. The top row on the right shows the results after individually removing the four most important layers, while the bottom row shows the results after removing the four least important layers. Removing highly important layers significantly degrades generation quality, whereas removing low-importance layers has minimal visual impact. The visual contrast provides strong evidence that layers contribute unequally to model performance, validating the effectiveness of importance-aware layer pruning.}
  \label{G2}
\end{figure}

\noindent\textbf{Layer Importance Analysis}.
In this part, we provide analysis of layer importance and the impact of different layers to the generation results.
The method for computing block importance is detailed in Section~\ref{Sandwich Training}. 
Figure \ref{G1} shows the importance scores of all blocks ranked from low to high.
It can be observed that only a few layers have importance scores significantly higher than others. 
This means that most blocks have a relatively small impact on the final generation, indicating room for optimization. Figure \ref{G2} shows a comparison of image generation results after removing layers of different importance. 
It can be observed that when important layers are removed, the model fails to generate images properly, while the generation results are almost unaffected when less important layers are removed. 
This validates the rationale behind our determination of layer importance.

\noindent\textbf{Visual Analysis of Multiple Replacement Strategies}.
To better understand the visual impact of different block replacement strategies, we present a qualitative comparison across four prompts and four methods in Figure \ref{compare}. Each row corresponds to a prompt, and each column represents a different approach: the baseline model without any modification, direct deletion of selected blocks, block-wise linear replacement without ST, and our proposed method combining BRLL and ST.
Visual results show that our method best preserves the original image characteristics. Direct deletion of blocks frequently results in noticeable texture degradation or missing structural elements, suggesting that individual blocks have varying levels of importance in preserving generation quality. Replacing blocks with linear layers without applying ST results in images that exhibit redundant patterns or positional artifacts, highlighting the limitations of naive replacement. In comparison, our full approach, which incorporates both BRLL and ST, produces images that are visually closest to the baseline. This demonstrates that BRLL provides a viable lightweight substitute for full blocks, and that ST is essential for stabilizing the training and aligning the replacement layers with the original model distribution. Together, these components effectively enhance inference efficiency while minimizing quality degradation.

\section{Conclusion}
In this work, we propose FastFLUX, a novel architecture-level pruning framework for efficient compression of large-scale diffusion models. By applying Block-wise Replacement with Linear Layers and introducing Sandwich Training for localized supervision, FastFLUX reduces inference cost and training overhead while maintaining high image quality. Even with 20\% of the hierarchy pruned, it remains robust in generation and structural consistency. Extensive experiments demonstrated FastFLUX’s effectiveness and generalizability across different models, offering a streamlined and flexible paradigm for efficient compression and acceleration of T2I diffusion models.

\newpage
\small
\bibliographystyle{unsrt}
\bibliography{references}

\end{document}